\documentclass[conference]{IEEEtran}
\IEEEoverridecommandlockouts
\usepackage{cite}
\usepackage{amsmath,amssymb,amsfonts}
\usepackage{algorithmic}
\usepackage{graphicx}
\usepackage{textcomp}
\usepackage{xcolor}
\usepackage{tikz}
\usepackage{url}
\usepackage{booktabs}
\usepackage{multirow}
\usepackage{multicol}
\usepackage{subcaption}
\usepackage{float}
\usepackage{titlesec}
\usepackage{enumitem}
\usepackage{dsfont}
\def\BibTeX{{\rm B\kern-.05em{\sc i\kern-.025em b}\kern-.08em
    T\kern-.1667em\lower.7ex\hbox{E}\kern-.125emX}}
\def\mathbi#1{\textbf{\em #1}}
\begin{document}
\abovedisplayskip=3pt
\abovedisplayshortskip=0pt plus 2pt
\belowdisplayskip=3pt
\belowdisplayshortskip=0pt plus 2pt

\title{Alioth: A Machine Learning Based Interference-Aware Performance Monitor for Multi-Tenancy Applications in Public Cloud
}

\author{\IEEEauthorblockN{Tianyao Shi\IEEEauthorrefmark{2}, Yingxuan Yang\IEEEauthorrefmark{2}, Yunlong Cheng\IEEEauthorrefmark{2}, Xiaofeng Gao\IEEEauthorrefmark{2}$^{*}$ \thanks{T. Shi, Y. Yang, Y. Cheng and X. Gao are with MoE Key Lab of Artificial Intelligence, Department of Computer Science and Engineering, Shanghai Jiao Tong University. X. Gao is the corresponding author.}, Zhen Fang\IEEEauthorrefmark{3}, Yongqiang Yang\IEEEauthorrefmark{3}}
\IEEEauthorblockA{\IEEEauthorrefmark{2}
 Shanghai Jiao Tong University, Shanghai, China}
\IEEEauthorblockA{\{sthowling,zoeyyx,aweftr\}@sjtu.edu.cn, gao-xf@cs.sjtu.edu.cn}
\IEEEauthorblockA{
\IEEEauthorrefmark{3}Alibaba Cloud, Hangzhou, China}
\IEEEauthorblockA{\{fangzhen8,yangyongqiang\}@alibaba.com}

\vspace{-2em}}

\maketitle

\begin{abstract}
Multi-tenancy in public clouds may lead to co-location interference on shared resources, which possibly results in performance degradation of cloud applications. 
Cloud providers want to know when such events happen and how serious the degradation is, to perform interference-aware migrations and alleviate the problem. 
However, virtual machines (VM) in Infrastructure-as-a-Service public clouds are black boxes to providers, where application-level performance information cannot be acquired. This makes performance monitoring intensely challenging as cloud providers can only rely on low-level metrics such as CPU usage and hardware counters.

We propose a novel machine learning framework, Alioth, to monitor the performance degradation of cloud applications. 
To feed the data-hungry models, we first elaborate interference generators and conduct  comprehensive co-location experiments on a testbed to build \emph{Alioth-dataset} which reflects the complexity and dynamicity in real-world scenarios. 
Then we construct Alioth by (1) augmenting features via recovering low-level metrics under no interference using denoising auto-encoders, (2) devising a transfer learning model based on domain adaptation neural network to make models generalize on test cases unseen in offline training, and (3) developing a SHAP explainer to automate feature selection and enhance model interpretability. 
 Experiments show that Alioth achieves an average mean absolute error of 5.29\% offline and 10.8\% when testing on applications unseen in the training stage, outperforming the baseline methods. Alioth is also robust in signaling quality-of-service violation under dynamicity. 
Finally, we demonstrate a possible application of Alioth's interpretability, providing insights to benefit the decision-making of cloud operators. The dataset and code of Alioth have been released on GitHub.
\end{abstract}


\section{Introduction}
Nowadays, cloud computing adopts multi-tenancy, i.e. instances of  multiple tenants sharing hardware resources on the same physical machine, to improve  resource utilization and cost efficiency~\cite{paragon}. Because of co-location, instances may contend on resources such as \emph{last-level cache (LLC), memory bandwidth, network bandwidth,} etc. \cite{bubble-up,ibench,ursa}. Resource contention may lead to interference, resulting in performance degradation of as much as 200\% compared with running in-isolation~\cite{sc12}, and quality-of-service (QoS) violation of cloud applications~\cite{sc12-google}.
 Cloud providers are to handle this side-effect of multi-tenancy.
 \textbf{Good monitoring\textemdash fast detection of interference and accurate estimation for performance} makes the first step to alleviating performance degradation, with which cloud operators can perform throttling or interference-aware migrations. The ultimate goal is to improve overall system efficiency and customer satisfaction.

\begin{figure}[t]
    \centering
    \includegraphics[width=0.99\linewidth]{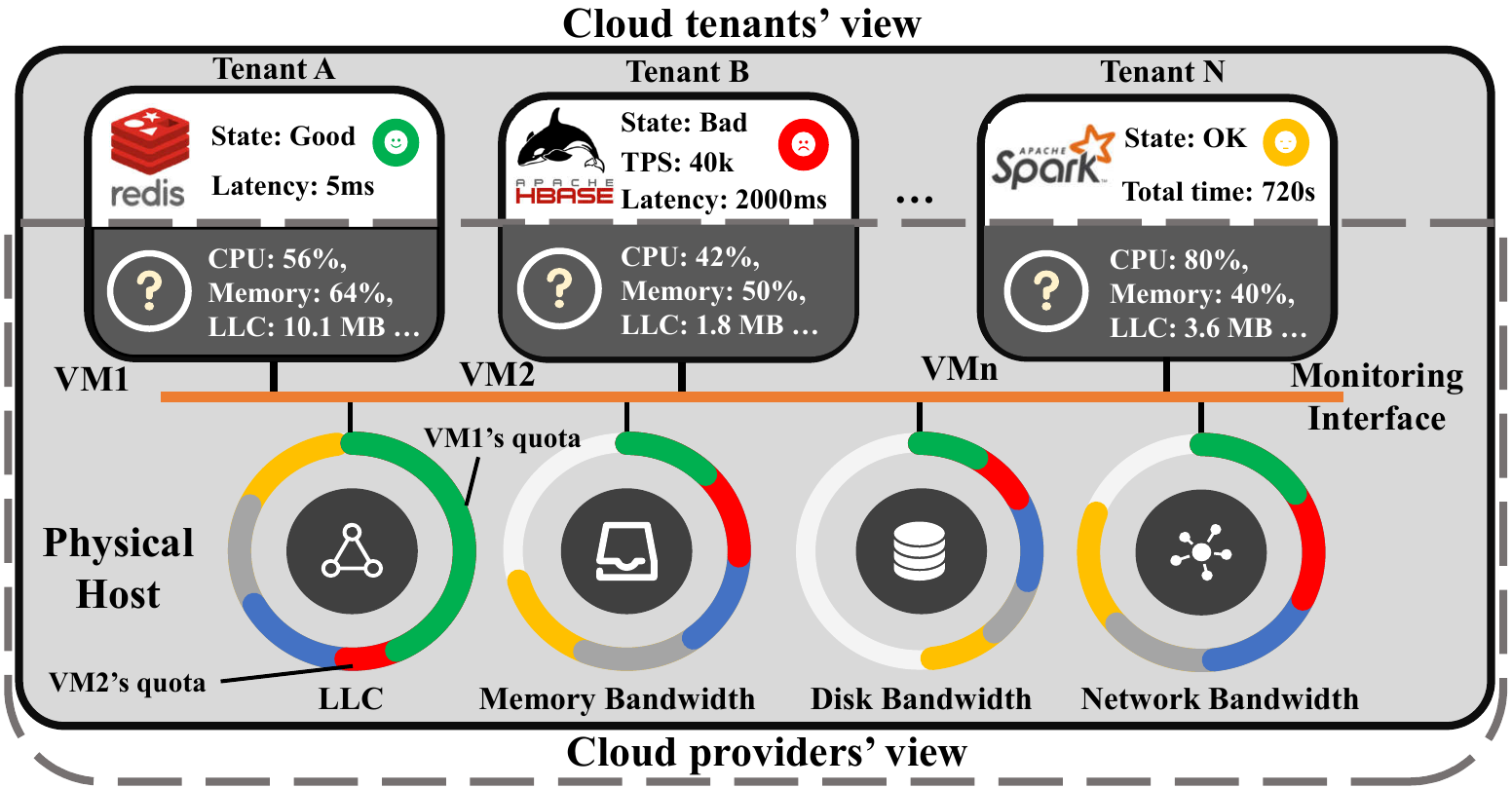}
    \captionsetup{font={footnotesize}}
    \caption{Multi-tenancy and black-box issue in public clouds. Multiple VMs share the same host and contend on shared resources. In this example, VM1 occupies too much LLC capacity such that VM2 cannot get the needed quota to run normally, resulting in severe performance degradation. Yet for cloud providers, VMs are black boxes and providers cannot acquire application-level information. They can only monitor low-level metrics such as LLC occupation, memory usage, etc. to estimate what is happening inside the VMs.}
    \label{fig:illustration}
\end{figure}

Unfortunately as Fig.~\ref{fig:illustration} depicts, virtual machines (VMs) are almost always \emph{black boxes} to cloud providers in public Infrastructure-as-a-Service (IaaS) cloud~\cite{rc,ml-centric}. Despite there being mechanisms like Application Heartbeats~\cite{heartbeats}, cloud customers are rarely willing to provide application-level performance information to the platform due to privacy issues~\cite{smartharvest}. Cloud providers can only rely on low-level metrics, e.g. CPU usage and hardware counters as proxies of performance. 
 The second challenge is the \emph{dynamicity} of cloud applications' workload intensity. 
Different workload intensity leads to different sensitivity and tolerance towards interference~\cite{bubble-flux}. Therefore, it may cause great errors if we use a static threshold for judging QoS violations under dynamicity. 
 The third challenge is that the relation between low-level metrics and performance is \emph{implicit}. Recent work~\cite{cpi-gbu} reports that the widely-used performance indicator, Clocks per Instruction (CPI) may be misleading when workload intensity changes.

    

Previous solutions to estimating performance degradation failed to address the three challenges completely, i.e.  \emph{black-box, dynamicity,} and \emph{impilicity}. Some assume application-level information available \cite{parties,bubble-up,proctor} does not fit into the black-box scenario in public clouds. Some introduce additional hardware \cite{asm} or online profiling that either becomes too costly \cite{deepdive,fecbench} or perturbs normal running \cite{paragon,bubble-flux,caliper}.  Some overly simplify the problem, only considering specific application category \cite{sc12,pythia}, limited cases of  co-location \cite{esp,jsa17}, static workload intensity \cite{sc12-google}, or use low-level metrics to directly represent application QoS \cite{cpi2,wsmeter}.

Machine learning (ML) methods are popular in existing literature \cite{fecbench,esp,monitorless}. The reason for using machine learning is mainly motivated by: (1) data of low-level metrics is abundant and easy to acquire under black-box scenarios; (2) the relationship between low-level metrics and  performance is implicit, and ML excels at discovering complex relationships between various factors. However, the results of ML models are often hard to interpret for humans. There may also be generalization problems as ML models often fail to work well when faced with unfamiliar feature distributions, which is likely to happen online when models encounter \emph{unknown} applications that are not in the training dataset.

We present a novel and interpretable ML-based framework, Alioth, as an interference-aware performance monitor for multi-tenancy applications in IaaS public clouds. 
Given the low-level metrics of a target VM, Alioth estimates how its performance changes compared with no interference. 
To feed the data-hungry ML models, we first devise a set of stressing programs to simulate various kinds of interference and conduct co-location experiments under various workloads and interference intensities on our testbed to make a dataset called \emph{Alioth-dataset}. 
We elaborate the experiments such that the considered cases can match real-world scenarios.

We then construct Alioth by the following steps to address the aforementioned challenges and limitations of ML methods: (1) We augment input features via recovering low-level metrics under no interference using denoising auto-encoders (DAE) \cite{DAE}. (2) To address the generalization issue in the online stage, we combine domain adaptation neural network (DANN) \cite{dann} and DAE to transfer knowledge learned in the offline stage and make the ML models generalize on \emph{unknown} applications. (3) We develop a SHAP \cite{shap} based explainer to automate feature selection and enhance model interpretability. 

Experiments show that Alioth not only achieves high accuracy\textemdash mean absolute error (MAE) of 5.29\% on average in estimating performance degradation offline but also maintains a decent average MAE of 10.8\% when testing with \emph{unknown} applications, outperforming the baseline methods, and is robust in signaling 0/1 QoS violation under dynamicity.
Finally, we use a case study to demonstrate how Alioth can perform attribution analysis, i.e., how much contribution each kind of interference makes towards current performance degradation. Thus, Alioth helps to gain more insights into the system by identifying the source of interference and benefits the decision-making of cloud operators.   

The dataset and code of Alioth have been made publicly available at Github\footnote{\url{ https://github.com/StHowling/Alioth}} to enhance reproducibility. In short, our main contributions are listed as follows:
\begin{itemize}
    \item We devise a set of interference generators and conduct co-location experiments to make a dataset that reflects the complexity of the real-world public cloud. 
    \item We present a novel ML-based framework as an interference-aware monitor of multi-tenancy applications in IaaS public clouds, outperforming previous ML-based solutions.
    \item We show that our design can enhance model generalization and interpretability.
\end{itemize}

The rest of the paper is organized as follows: Section~\ref{sec:rw} discusses related work. Section~\ref{sec:model} formalizes the problem and presents data observations. Section~\ref{sec:method} describes the methodology for creating an accurate and explainable ML-based model. Section~\ref{evaluation} evaluates the accuracy of the model. Section~\ref{sec:conclusion} concludes our works.
  \section{Related Works}\label{sec:rw}
There has been a large amount of work that focuses on predicting performance degradation of cloud applications caused by shared resource interference \cite{asm,caliper,deepdive,sc12,esp,monitorless}. According to their methodologies, previous works for predicting QoS degradation can be divided into non-ML methods and ML methods. 
\subsection{Non-ML Methods}
Non-ML methods can be further categorized as follows:\par
\noindent\textbf{Focusing on white-box scenarios:} Works in this category assume that application-level performance can be acquired via some interface.
PARTIES \cite{parties} and Heracles \cite{heracles}  guarantee QoS for latency-sensitive applications via dynamic feedback control. Proctor \cite{proctor} uses a median filtering algorithm and step detection to look for abrupt changes in the QoS. 


\noindent\textbf{Using classical performance indicators:} Works that follow this category assume a linear relation between QoS and metrics like CPI.  Bubble-up \cite{bubble-up}, $\text{CPI}^2$ \cite{cpi2} and Wsmeter \cite{wsmeter} use CPI directly as ground-truth performance indicator. Recent research \cite{cpi-gbu} shows that CPI could be misleading as shifts in workload intensity may cause significant changes in CPI. 


\noindent\textbf{Using additional hardware or online profiling:} Deepdive \cite{deepdive} clones the target VM to profile it online. Application Slowdown Model (ASM) \cite{asm} devises non-standard hardware counters to calculate slowdown caused by cache interference. Bubble-flux \cite{bubble-flux} and Caliper \cite{caliper}  periodically hang up all other VMs except the target to get in-isolation performance. However,  hanging up an interactive application could cause catastrophic QoS violations.	

Alioth differs from the works above in that: Alioth (1) assumes no application-level information available, (2) models the relationship between performance and low-level metrics in a complex way, and (3) does not require costly additional hardware or online profiling. Alioth also considers dynamicity by incorporating resource usage metrics in its inputs.

\subsection{ML Methods}
There is also a large amount of work in recent years that use ML models such as linear regression (LR), regression trees (RT), support vector (SV) regressions, and neural networks (NN) to predict performance degradation. 
In 2012, Dwyer et al. \cite{sc12} first proposed Practical Method, using machine learning for modeling performance degradation of HPC workloads.
They tried a variety of ML models and found that bagged RT performs best on their dataset. 
Later many works propose LR-based models \cite{micpro16,pythia,fgcs20} to predict the degradation of total execution time for long-computing tasks. 
ESP \cite{esp} introduces a regularization method to perform feature selection. 
DIAL \cite{dail} uses a decision tree (DT) model and a queuing model. 
Cheng et al. \cite{jsa17} use ensembled LR, RT, and NN models to predict performance degradation of pairwise co-located VMs.
There is also literature that involves performance prediction but does not estimate  the performance degradation of a current running VM. Monitorless \cite{monitorless} trains binary classifiers for identifying resource saturation in components of microservices. 

Alioth goes beyond the ML-based methods mentioned above with an elaborated design. We augment input features with denoising auto-encoders and develop a SHAP explainer to enhance interpretability, providing insights to benefit the decision-making of cloud operations.
 \section{Modeling, Data, and Observations}\label{sec:model}
In this section, we first model the problem of estimating applications' performance degradation in public clouds, describe the nomenclature, the input and target of the ML model, and how the ground-truth label of degradation is defined. Then  we show how to build \emph{Alioth-dataset}, a dataset that well represents the variety of application categories, workload intensity, and interference type and intensity in the public cloud. Finally, we present some key observations from \emph{Alioth-dataset} that motivate the design of ML models used in Alioth.

\subsection{Modeling}
A \emph{cloud} is a set of physical nodes $\mathcal{P} =\{p_1,\cdots,p_M\}$, where $M$ is the total number of nodes. Each node $p \in \mathcal{P}$ can host many instances as VMs or Linux containers in a virtualized environment. For simplicity, we only consider VMs in Alioth, but the methodology can be seamlessly transferred to suit containers. Let $v_{i,j}$ denote the $i$-th VM on host $p_j$, and $\mathcal{V}_j =\{v_{1,j},\cdots,v_{n_j,j}\}$ be all $n_j$ VMs that are currently running and sharing resource on $p_j$. Each VM may be running one or several applications or utilities, and some VMs may work together to form a high-level multi-tier cloud service. We only consider the performance degradation of every single VM due to co-location interference. 

At each timestamp $t$, the cloud provider is able to collect a set of \textbf{host metrics} $R_{p,t}$ which describe the overall resource usage of node $p$ (e.g., memory capacity and bandwidth usage of the whole CPU socket), as well as two sets of \textbf{VM metrics} $R_{v,t}$ and $H_{v,t}$, which reflects resource usage (e.g., CPU/memory capacity, disk, and network packet read/write) and hardware events (e.g., cache miss, branch mispredict) of VM $v$ respectively. For all VMs $v \in \mathcal{V}_j$ that are hosted on the same node $p_j$, they share the same value on the overall host resource usage $R_{p_j,t}$. Moreover, there is a metric $\mathcal{Q}_{v,t}$ that represents the current performance of the application(s) running in $v$, e.g. response latency, throughput, or finishing time, but it cannot be observed by the provider online.

Suppose $T^{*}$ denotes the set of timestamps when there is no co-location interference, and $\overline{\mathcal{Q}}_{v,T^{*}}^{(i)}$ refers to the average performance when there is no interference, then the performance deviation $d_{v,t}^{(i)}$ of VM $v$ running with workload intensity $i$ at timestamp $t$ is defined as
$$d_{v,t}^{(i)} =\mathcal{Q}_{v,t}^{(i)} / \overline{\mathcal{Q}}_{v,T^{*}}^{(i)} = \mathcal{Q}_{v,t}^{(i)} /  \frac{\sum_{t \in T^{*}}\mathcal{Q}_{v,t}^{(i)}}{|T^{*}|}.  $$
If smaller $\mathcal{Q}_{v,t}^{(i)}$ means better performance (e.g. latency), the \textbf{performance degradation} $D_{v,t}^{(i)}$ is defined as $d_{v,t}^{(i)}-1$. Otherwise, it is defined as $1-d_{v,t}^{(i)}$.

The problem of estimating performance degradation of some VM $v$ at timestamp $t$ is then stated as follows: 
given low-level metrics $\mathcal{M}_{v,t}=(R_{v,t},H_{v,t},R_{p,t})$, 
the model should give an estimation of degradation $\hat{D}_{v,t}$. 
For the machine learning problem, a dataset $\mathcal{D}=\{(\mathbi{x}_{v,t},y_{v,t})\}$
is built from VMs running different applications under various workload intensities and co-location schemes to mimic the complexity of real-world public cloud interference,
where $\mathbi{x}_{v,t}$ is the feature vector built upon $\mathcal{M}_{v,t}$ and $y_{v,t}=D_{v,t}$.
The goal of ML is to learn a function $f$ mapping $\mathbi{x}_{v,t}$ to $\hat{y}_{v,t}=\hat{D}_{v,t}$ such that the difference between $\hat{y}_{v,t}$ and $y_{v,t}$ is minimized.

\subsection{Building Alioth-Dataset}\label{sec:3-1}
\noindent\textbf{Experimental Setup:}
Table~\ref{tab:hardware} lists the major hardware specifications of the experimental platform. We conduct our experiments on 7 such physical nodes, each equipped with 500G memory and connected by a 10Gbps switch. This testbed hosts up to 100 VMs simultaneously, where each VM runs one specific application and is allocated 4 logic cores and 8GB memory by default. VMs are allocated the same amount of computing resources for simplicity, but our framework supports varied flavors of VM scales. Due to the time limit, we did not enable CPU over-committing to simulate extremely crowded co-location scenarios when collecting data, and we leave this issue for future work.
\begin{table}[htbp]
    \centering
    \setlength{\abovecaptionskip}{0.1cm}
    \caption{Hardware Specification}
    \label{tab:hardware}
    \begin{tabular}{c c}
    \toprule
        \textbf{CPU} & Intel Xeon Gold 6151 \\
        \textbf{Virtualization technology} & QEMU-KVM + Openstack \\
        \textbf{Host OS} & Red Hat 2.8.5 (kernel 3.10)\\
        \textbf{Guest OS} & Ubuntu 16.04 (kernel 4.14) \\
        \textbf{Cores, Sockets} & 2 sockets, 18 physical cores per socket \\
        \textbf{Core frequency} & fixed at 3.0GHz, DVFS disabled  \\
        \textbf{Hyperthreading} & Enabled, 2 threads per physical core \\
        \textbf{Last-level (L3) cache} & 22.5MB, 11 ways \\
    \bottomrule
    \end{tabular}

\end{table}

\noindent\textbf{Application Selection:}
The following describes the cloud applications we use to build \emph{Alioth-dataset}, along with the corresponding method to generate their workloads. The applications can be roughly categorized into the following 4 classes: 
(1) \textbf{NoSQL databases}: We use \emph{Cassandra} and \emph{HBase} (wide-column store), and \emph{MongoDB} (document store) as representative NoSQL applications, and choose YCSB \cite{ycsb} as a unified workload generator for the three applications. We change read/write ratios, request distributions, and query-per-second (QPS) to simulate real-world workload flows. 
(2) \textbf{Message middlewares}: They are vital building blocks of large-scale cloud services nowadays. We choose \emph{Kafka} and \emph{RabbitMQ} and use their official benchmark tools as workload generators. We change the number of clients and QPS to achieve dynamic workload intensity. 
(3) \textbf{Key-value stores}: They serve as a cache between disk-bound databases and main memory. We use \emph{Redis} and \emph{Etcd} and use their official benchmark tools as workload generators.
(4) High-performance computing (\textbf{HPC}): We use applications in \emph{SPEC CPU 2006} benchmark suite as HPC workloads, and adjust its intensity via changing the threads number running in parallel.
Alioth is open to other kinds of cloud applications that are not mentioned above, such as web servers, search engines, big data analytics, ML inference, and multi-tier services. We are confident in building ML models that perform well on VMs that run them, either by collecting their labeled data and retraining the models or by adapting the models to the unlabelled data that is distributed differently with \emph{Alioth-dataset} via transfer learning.

\noindent\textbf{Generating Interference:}
We use two different ways to generate interference on the shared resources. The first is to implement specialized stressing \emph{Bubbles}, or micro-benchmarks \cite{bubble-up,ibench}, which are adjustable small programs that exhaustively thrash target shared resources to interfere with co-located VMs. We consider four kinds of shared resources that may cause interference: \textbf{LLC}, memory bandwidth (\textbf{MBW}), network bandwidth (\textbf{NBW}), and disk bandwidth (\textbf{DBW}). 

Here we briefly describe the implementation method of the four bubbles. (1) For LLC, the bubble continuously reads chunks of memory where each chunk equals the size of a cache line to occupy LLC space. We use \emph{Intel-CAT} \cite{intel-cmt-cat} to restrict the number of cache ways the bubble can use to adjust interference intensity: when the bubble can use all cache ways, the induced interference is the strongest. (2) For MBW, the bubble continuously copies a fixed chunk of memory that equals cache line size, such that each operation triggers one cache miss and memory access. The bubble is constrained to use only 1 LLC cache way for 1 thread to avoid further influence on LLC usage. The interference intensity is controlled by parallel thread numbers. (3) For NBW, the bubble uses \emph{iperf3} \cite{iperf3} to send UDP packets to another VM as a client located on another host. We adjust the consumed bandwidth by configuring \emph{iperf3} parameters. The largest interference a single bubble can issue is 1500Mbps when the client starts to drop packets. (4) For DBW, the bubble uses \emph{fio} \cite{fio} to perform asynchronous reads and writes. By changing queue depths and the number of parallel threads, the bubble adjusts the I/O operation per second.

Another way to generate interference is to co-locate the applications on the same host directly. We set up 2 to 5 VMs on the same host socket to simulate real-world co-location interference, where each VM runs one of 8 considered applications at a certain workload intensity. That is, a VM may or may not be running the same application with other co-located VMs, but probably at a different intensity.  We try 35 different co-location schemes and an average of 5 different representative workload intensities for each of the eight applications mentioned above. The details of the co-location schemes can be found in \emph{Alioth-dataset}.

\begin{figure}[ht]
    \centering
    \subcaptionbox{\label{fig:stressor}}{\includegraphics[width=0.49\linewidth]{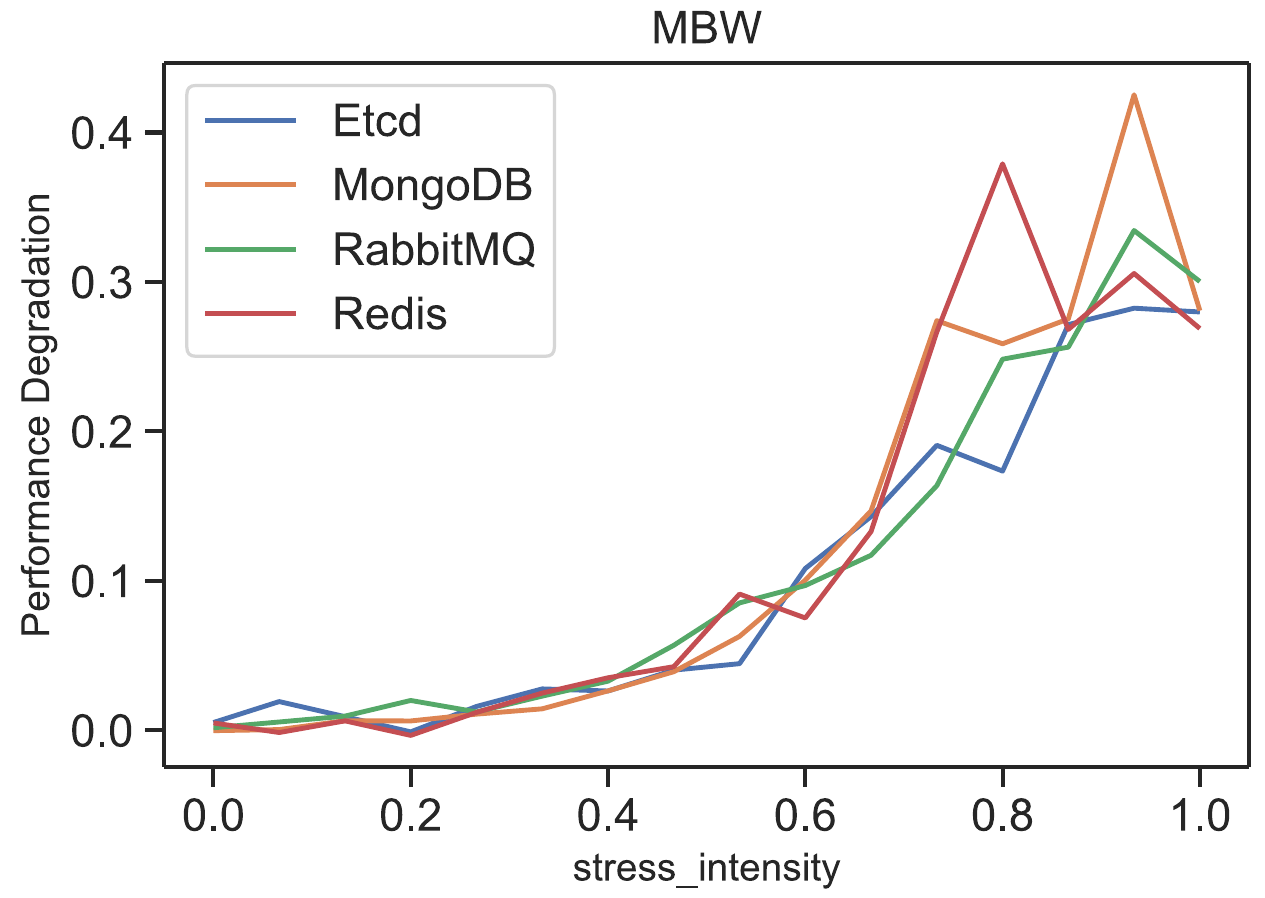}}
    \subcaptionbox{\label{fig:app-varstress}}{\includegraphics[width=0.49\linewidth]{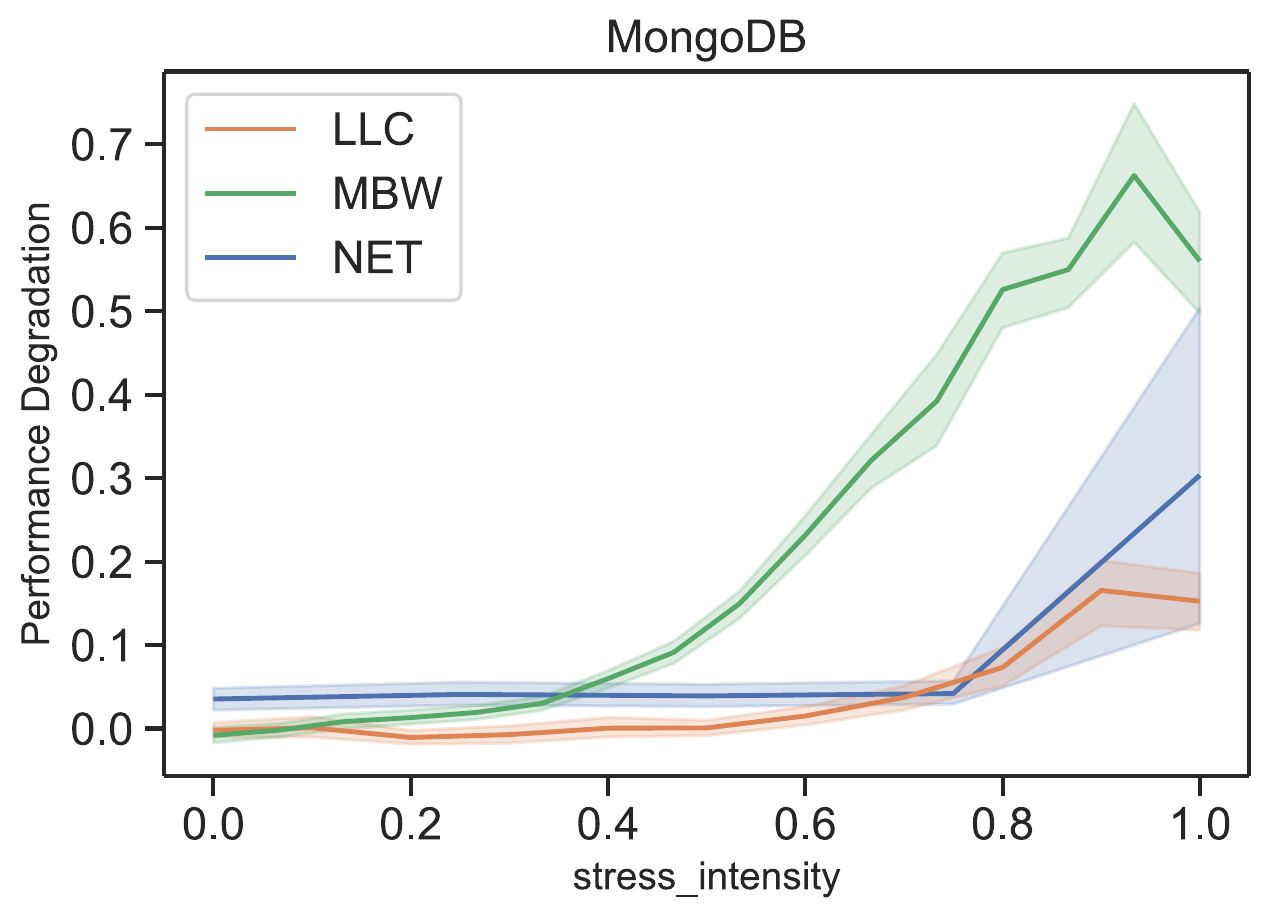}}
    \setlength{\abovecaptionskip}{0cm}
    \caption{(a) Generating MBW interference with adjustable intensity leads to significant performance degradation of  applications. The interference intensity and performance degradation are approximately positively linear-correlated.
    (b) MongoDB as an example to show the effectiveness of bubbles. As the interference intensity increases, all bubbles gradually start to deteriorate application performance.
    }
\end{figure}

Figure~\ref{fig:stressor} and~\ref{fig:app-varstress} shows the effectiveness of the interference generation process discussed above. Note that 4 other applications and the DBW bubble are omitted from the figures for beauty considerations, as their lines largely overlap with others. The DBW bubble's behavior is very like that of NBW since the disks in today's cloud lie in a remote storage pool and are attached through network connections.

\noindent\textbf{Comments on Choice of Selection:} We consider these shared resources and applications as they are typical in previous literature, e.g., Paragon \cite{paragon}, PARTIES \cite{parties}, URSA \cite{ursa}, and Alita \cite{alita}. Although cloud services nowadays seldom run as a standalone application in a single container and are more often organized as collections of micro-services, each micro-service component operating in a separate container is still suitable for the methodology of data collection described above.

Although we have tried our best to simulate real-world scenarios, data collected into \emph{Alioth-dataset} is still a small subset of all possible co-location interference cases. This is due to the exponential scale of all configurations w.r.t. applications, number of co-located VMs, interference type and intensities, workload type and intensities, hardware specifications, etc. Therefore, building a dataset that includes all possibilities is simply intractable. When the models are tested online, they are bound to encounter unseen and differently distributed data compared to \emph{Alioth-dataset}. We must bear this in mind when designing ML models.

\noindent\textbf{Collecting Metrics:}
We collect the low-level metrics $\mathcal{M}_{v,t}=(R_{v,t},H_{v,t},R_{p,t})$ separately. For resource usage $R_{v,t}$ of VM $v$, we use \emph{libvirt}; for hardware events $H_{v,t}$, we use Linux \emph{perf}, and for overall resource usage $R_{p,t}$ of host $p$, we use Linux \emph{sar} and \emph{Intel-pqos}. These metrics are monitored and collected in the physical host OS. The performance metric $\mathcal{Q}_{v,t}$ is collected by the workload generator in client VMs that are allocated 16 physical cores with 32G memory each and located on a specialized host node to prevent the client from being the performance bottleneck. For HPC applications, we record their CPI; otherwise, we record both the throughput and latency. Although execution time as the performance label is more common for HPC applications, there are also papers using CPI (e.g., Caliper \cite{caliper}) as a performance label. CPI is more convenient for data collection, but we are planning to add execution time as labels in the future.

\noindent\textbf{\emph{Alioth-dataset} Overview:}
Merging all collected data, we get around 46k data samples to make \emph{Alioth-dataset}, where each sample is a tuple of $(\mathcal{M}_{v,t}, \mathcal{Q}_{v,t}, \mathcal{A}_{v,t})$ in a one-second period, and  $\mathcal{A}_{v,t}$ is the additional label describing the application, workload intensity, interference type, and interference intensity of VM $v$ at timestamp $t$. Each sample has 224 columns, among which 15 belong to $R_{v,t}$, 175 belong to $H_{v,t}$, 28 belong to $R_{p,t}$, and the rest are performance labels and additional labels. 

\subsection{Key Data Observations}\label{sec:observation}
In this part, we do some preliminary study into \emph{Alioth-dataset}, exploring in detail the  \emph{implicity} problem when we directly use certain low-level metrics $m_{v,t}\subset \mathcal{M}_{v,t}$ for estimating performance.

\noindent\textbf{Data Preprocessing:}
We first delete the features whose values and variances are all 0 since they could not provide any effective information. Then we omit extreme values in the raw data that may appear due to hardware counter overflow or inaccuracy from multiplexing and sampling. After that, we use MinMaxScaler to normalize the remaining metrics, making the features fall in the same numerical magnitude of [0,1].

\noindent\textbf{Exploring \emph{implicity}:}
Once we determine the workload intensity of a VM, we can then examine the accuracy of directly using deviations $\hat{d}_{v,t}^{(i)}$ of selected low-level metrics $m_{v,t}^{(i)}$ to represent performance degradation. We consider 2 different approaches: (1) \emph{Best-Possible}: Current workload intensity $i$ and corresponding performance baseline $\overline{m}_v^{(i)}$ are already known, which is impractical in reality. (2) \emph{Best-Effort}: We estimate $i$ and $\overline{m}_v^{(i)}$ from historical observations by fitting two Gaussian mixture models (GMM)~\cite{gmm}. The first fitting operates on memory usage to find the historical samples with the same workload intensity $i$. The second operates on $m_{v,t}^{(i)}$ of these samples to find the cluster with the lowest (highest) mean, and use the mean value as the performance baseline.

\vspace{-0.1cm}
 \begin{figure}[ht]
    \centering
     
    \includegraphics[width=0.8\linewidth]{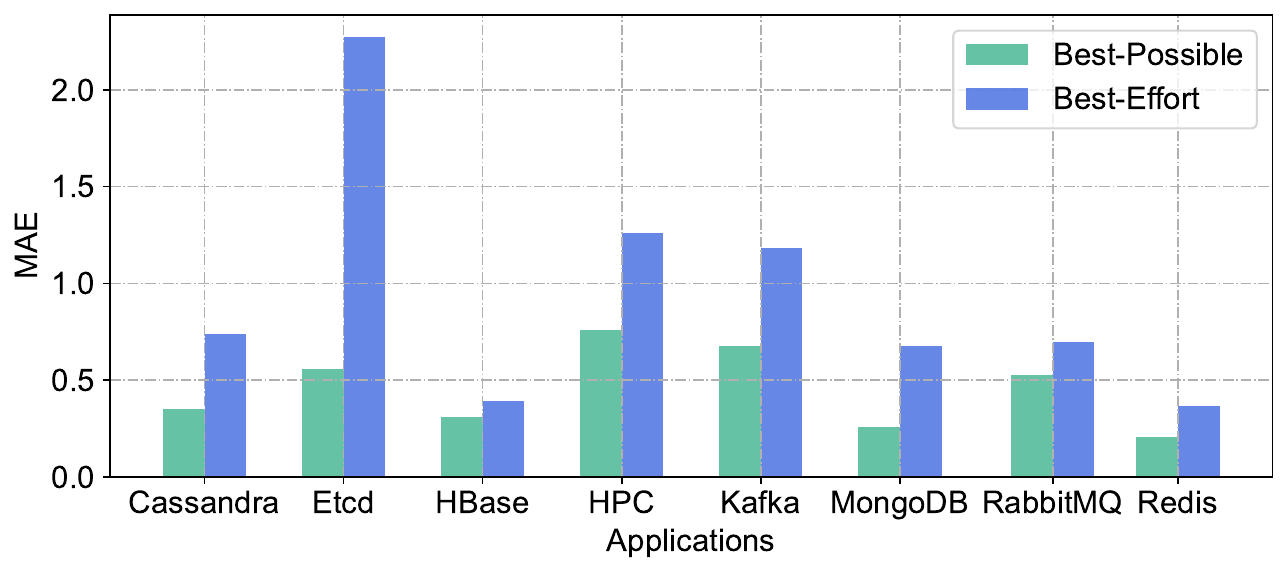}
    
    \caption{Accuracy of using CPI under 2 approaches to estimate performance.}\label{fig:CPIBestWorst}
\end{figure}

Here we only present the result of CPI, for (1) CPI is the most frequently used low-level performance metric in previous literature \cite{cpi2,sc12-google}; (2) we tried other metrics such as cache misses per kilo instructions (MPKI), and the error can be as high as 5.0x of CPI's result. Figure~\ref{fig:CPIBestWorst} compares the 2 approaches and the result suggests that directly using CPI can lead to great error even for \emph{Best-Possible} with an MAE of 45.7\%, and the \emph{Best-Effort} method makes things worse. It is only when the QoS violation threshold is loose can these methods be acceptable, and they are only suitable for signaling 0/1 QoS violation, rather than providing precise estimation for performance degradation.


\begin{figure*}[t]
    \centering
    \setlength{\abovecaptionskip}{0cm}
    \setlength{\belowcaptionskip}{-0.4cm}
    \includegraphics[width=\linewidth]{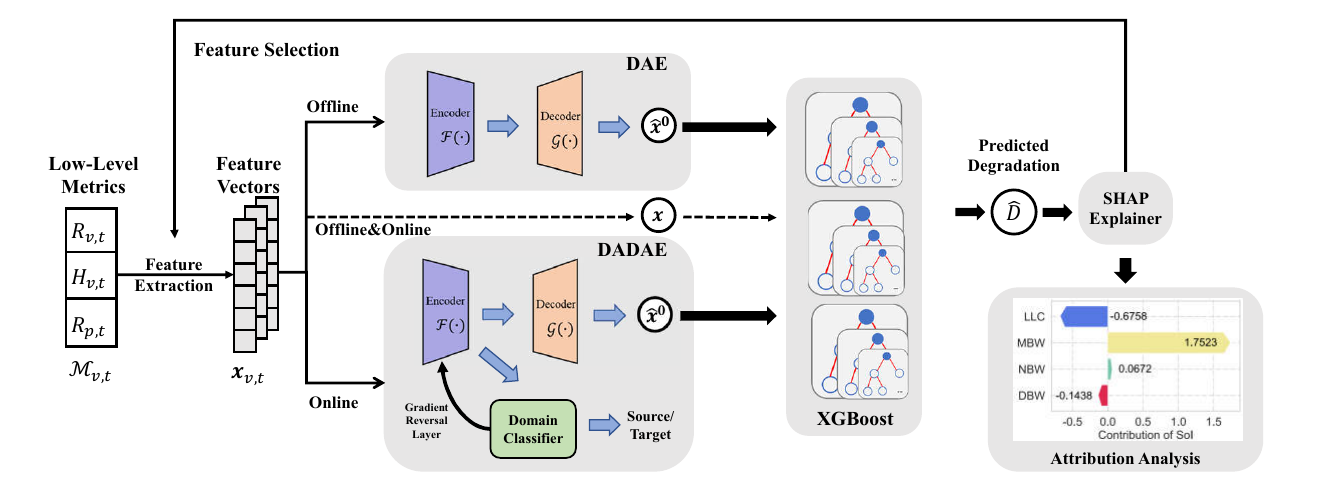}
    \captionsetup{font={small}}
    \caption{System design of Alioth. Alioth has three major components: an application classifier, a performance estimator, and a SHAP model explainer to make the system a holistic and interpretable approach to monitoring application performance.}
    \label{fig:alioth}
\end{figure*}




\section{Methodology of Alioth}\label{sec:method}
In this section, we first present the framework of Alioth and an overview of its design.  Then we move to the critical techniques we use in constructing Alioth,  i.e. denoising auto-encoder (DAE), domain adaptation neural network (DANN), eXtreme Gradient Boosting trees (XGBoost), and SHAP. We explain their principles and rationales in detail.
Finally, we present the feature selection and result attribution analysis process, as applications of the SHAP explainer.

\subsection{Design Overview}

Faced with the \emph{black-box, dynamicity, implicity} challenge,  and the lack of generalization and interpretability of previous ML methods, we present our design of Alioth in Figure~\ref{fig:alioth}. 
Alioth operates differently in the offline and online stage. Offline Alioth consists of 3 major parts: \textbf{a data augmenter based on DAE, a XGBoost performance estimator, and a SHAP model explainer}, each with different functions. Online Alioth replaces the DAE with \textbf{domain adaptation DAE (DADAE)}, i.e. the combination of DAE and DANN, to improve the model generalization. 

Given a target VM $v$, Alioth first extracts features $\mathbi{x}_v$ from low-level metrics $\mathcal{M}_v$ as the feature engineering phase. For each time stamp $t$ and each low-level metric, We extract a set of first-order statistics from time windows $[t-T,t]$ of various lengths $T$ ($t>T$). Specifically, the first-order statistics include the \emph{mean, minimum,} and \emph{maximum values}, the \emph{difference between max-min values}, and the \emph{standard deviation}. The time window length $T$ varies from 3, 5, 10, and 20 seconds.


As for the offline ML models, we use DAE \cite{DAE} and XGBoost \cite{xgb}.
A key data observation we find is that interference can be viewed as noise applied on low-level metrics under no interference to synthesize interfered features.
DAE is used to augment the input features via denoising $\mathbi{x}_{v,t}$ and recovering possible $\hat{\mathbi{x}}_{v,t^{ni}}$.
The two forms of features are merged for XGBoost to compare and estimate current performance. 


The online DADAE model combines DAE with DANN \cite{dann}. After the models are trained offline and before they are pushed online, we first collect some unlabelled data from the online environment to act as the road map of transfer learning. The unlabelled online data and labelled offline data are fed into DADAE together. With the help of the domain classifier, we conduct adversarial training to transfer the denoising encoding ability of DAE encoder, such that the tuned DADAE can output accurate result on unseen data. 

Finally, we develop a model explainer based on SHAP to perform feature selection and enhance model interpretability. It automates the feature selection process and provides insights for cloud operators to make optimizing decisions.

We implement components in Alioth using Python 3.8.0, \emph{scikit-learn} 0.24.1, \emph{Pytorch} 1.8.0 with \emph{CUDA} 10.2, and \emph{xgboost} 1.5.0. The model is trained on an NVIDIA V100 GPU, but does not necessarily require so much computing resource to be trained well. As we will see in Sec. V-B, Alioth uses fairly lightweight models that converge and infer fast.

\subsection{ML Based Models}\label{sec:mlmethods}
Here we briefly demonstrate the critical techniques we use in designing Alioth and the motivation for choosing them.

\subsubsection{Denoising Auto-Encoder: }
We use DAE to augment the input features by providing what the features should be like under no interference with their current values to the regressor $r$, such that the performance change can be inferred by comparing the difference between them. This design is based on the following evidence observed from data: the values of low-level metrics can be seen as a combination of interference noise and no-interference values.


 \begin{figure}[t]
    \begin{minipage}[t]{0.33\linewidth}
    \vspace*{\fill}
        \includegraphics[width=\linewidth]{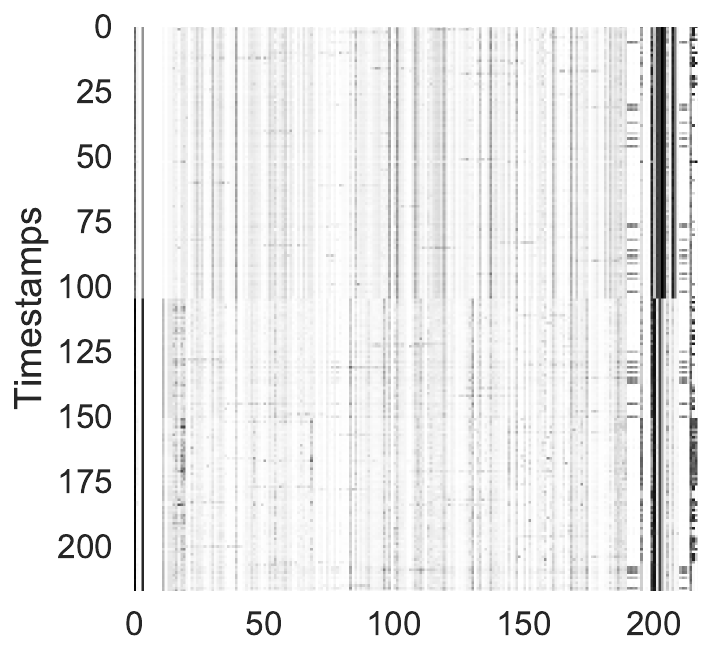}
        \parbox[][1pt][c]{\linewidth}
        \subcaption{}
        
    \end{minipage}
    \begin{minipage}[t]{0.31\linewidth}
    \vspace*{\fill}
        \includegraphics[width=\linewidth]{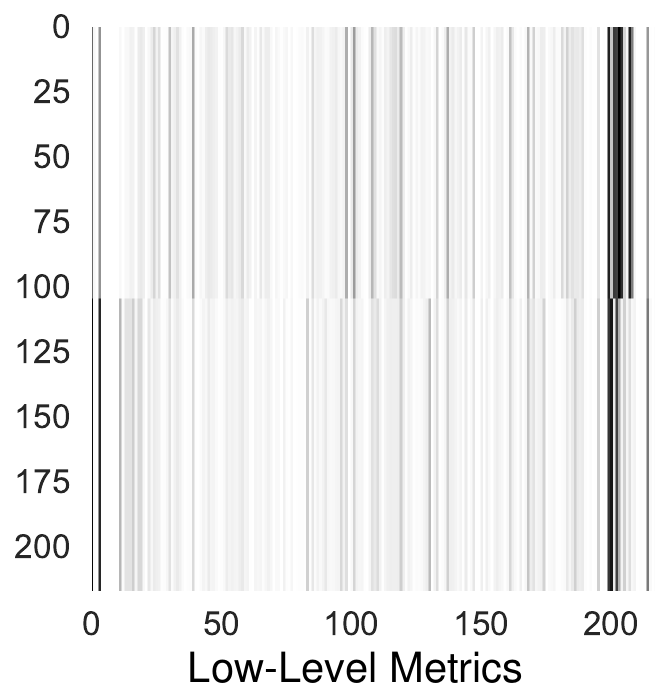}
        \subcaption{}
    \end{minipage}
    \begin{minipage}[t]{0.31\linewidth}
    \vspace*{\fill}
        \includegraphics[width=\linewidth]{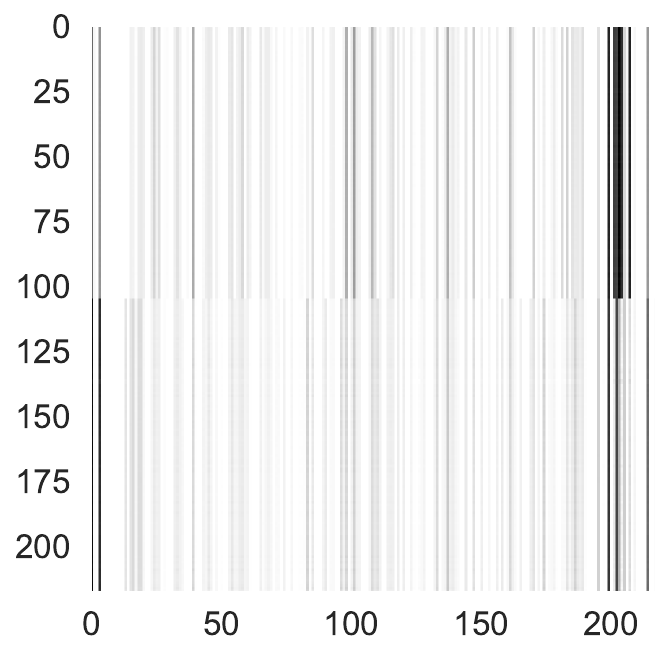}
        \parbox[][1pt][c]{\linewidth}
        \subcaption{}
    \end{minipage}
    \setlength{\abovecaptionskip}{0cm}
    \setlength{\belowcaptionskip}{0cm}
    \caption{Comparison of Cassandra's (a) interfered data, (b) data under no interference and (c) the interfered data processed by DAE. The horizontal axis represents 218 low-level metrics. The vertical axis represents 220 consecutive timestamps.}\label{fig:DAE}
\end{figure}

Figure~\ref{fig:DAE} shows three grayscale data plots for 218 metrics at 220 timestamps for a VM running Cassandra. The data in (a) corresponds to when it is being interfered with, which is the input of DAE. The data in (b) has the same workload intensity as that of (a) but is collected without interference and acts as the target of DAE in training. The difference in data distribution before and after timestamp 100 is due to a shift in workload intensity. We can see that there are glitches and perturbations in (a) compared with that of (b), which depicts the interference effects as \emph{noise} applied on metrics under no interference.  Data in (c) is the output of DAE and shows that DAE can denoise the interfered data, approximating metrics under no stress. Data processed by DAE slightly deviate from data under no interference, meaning that DAE does not work by memorizing.

\subsubsection{Domain Adaptation DAE}
We have mentioned the problem of unseen and differently distributed data. To demonstrate its severeness, we train DAE and XGBoost on the training set which excludes an application $X$, and test the models on the testing set that solely contains data of $X$. Compared to the standard setting where both training and testing set contains $X$ (see Tab.~\ref{tab:online}), the average error reaches \textbf{0.4597}, about \textbf{8x} of normal offline performance. The most extreme case happens to Etcd, where in normal case the error is \textbf{0.0277}, while in this setting the error is \textbf{1.5466}, reflecting very different data distributions between Etcd and the rest applications.

As said in Sec. III-B, the models are bound to face unseen cases not in the training set when running online, for which the ML models that cannot generalize on unseen and differently distributed data are impractical. As XGBoost has shown good generalization through its wide industrial use, transferring the denoising ability of DAE to unseen data becomes critical. Therefore, we combine the classical transfer learning model DANN with DAE to devise DADAE.

Specifically, DADAE contains three components: (1) an encoder that maps the feature vector $\mathbi{x}_{v,t}$ to a new feature space, where not only is the invariant information representing the no-interference states is maintained, but also the data from source domain (i.e. data seen in offline training stage) and target domain (i.e. unlabelled online data) is confused. In other words, the encoder tries to learn a domain-invariant feature representation. (2) A decoder that decodes the representations learned by the encoder into feature vectors under no interference is the same as the original DAE. (3) A domain classifier that helps the encoder to achieve its goal via adversarial training. Note that the Gradient Reversal Layer (GRL) \cite{dann} is used to simplify training. 

The optimization target of DADAE can be described as
\begin{equation*}
\resizebox{\linewidth}{!}{
$E(\theta_{\mathcal{F}},\theta_{\mathcal{G}},\theta_{\mathcal{D}})=\sum\limits_{\substack{i=1,\cdots,N \\ d_i=0}}L_R^i(\theta_{\mathcal{F}},\theta_{\mathcal{G}})-\sum\limits_{i=1,\cdots,N}L_d^i(\theta_{\mathcal{F}},\theta_{\mathcal{D}};R_{\lambda})$,
}
\end{equation*}
where $L_R(\cdot,\cdot)$ is the reconstruction error of DAE (\emph{min-square loss}), and $L_d(\cdot,\cdot)$ is the loss of domain classification (\emph{logistic loss}), $d_i=0(1)$ refers to data from the source (target) domain, $R_{\lambda}$ is the special network layer GRL, and $L_R^i, L_d^i$ means the loss w.r.t. sample $i$. The optimization process can be done with standard stochastic gradient descent.
\subsubsection{XGBoost}

We use XGBoost \cite{xgb} as the regressors in the performance estimators for its high accuracy, strong robustness, good generalization, fast convergence and inference, and light-weight model size. For the training process of performance estimators, we first train the DAEs by minimizing the denoising loss, and then fix the DAEs to optimize XGBoost models.
To fully exploit the generalization ability of XGBoost, we use 5-fold cross validation and grid search to adjust the  parameters in a 3-pass process, sequentially tuning tree splitting, data augmentation and regularization parameters.

\subsubsection{SHAP}
SHAP \cite{shap} is a ``model interpretation'' package developed in Python which builds an Additive interpretation model inspired by cooperative game theory. For each predicted sample, SHAP leverages the model being interpreted to generate a predicted value, and outputs the Shapley value assigned to each feature. 
When using SHAP, we pass the trained ML model and the prediction results of the test set into the interpreter, which calculates and outputs the global and local Shapley values. Shapley value reflects the influence of features in each sample and shows the polarity of the influence.

\subsection{Feature Selection and Model Interpretability}
\subsubsection{Feature Selection}
Alioth first uses SHAP to select features. To evaluate the importance of different features, Alioth first uses the Wrapper method by SHAP to calculate the Shapley value of each feature in the prediction model. Then it sets an appropriate threshold value, and selects the indices with Shapley values above the threshold as selected features. Take Cassandra as an example, Table~\ref{tab:feature selection} shows the top 10 metrics ranked by absolute global Shapley value (i.e. importance), where \emph{MBR} and \emph{MBL} refer to memory bandwidth usage of remote and local sockets respectively; \emph{net\_rd\_byte} and \emph{net\_wr\_byte}  refer to network read and written in bytes. Cassandra is a NoSQL database whose workload pattern involves memory reads and network I/O. Therefore, changes in these metrics may indicate performance change, which is in line with engineers' intuition. We use the result of feature selection to retrain the models used in Alioth.

Another function of feature selection is to reduce the number of metrics to monitor in production, as collecting 224 features at system runtime is costly. We only select features that show general importance among most applications: for each application, we select the top 20 largest features in terms of positive and negative values of global importance (which means 40 features). We then select the 22 features appearing in the 40-feature set for more than 4 out of 8 applications in \emph{Alioth-dataset} as final features to use. The details can be found in the released dataset.

\begin{table}[t]
\caption{Top 10 features by importance for Cassandra.} 
\label{tab:feature selection}
\centering
\begin{tabular}{c}
\hline
MBR.10.maxidff                                  \\ MBL.10.maxdiff                                      \\
net\_rd\_byte.5.std                        \\ UNHALTED\_REFERENCE\_CYCLES.10.std              \\
kbcached.10.mean                             \\ system\_time.5.maxdiff                             \\
MEM\_INST\_RETIRED:STLB\_MISS\_LOADS.10.mean \\ net\_wr\_byte.5.std                            \\
RS\_EVENTS:EMPTY\_CYCLES.10.std             \\ CORE\_SNOOP\_RESPONSE:RSP\_IFWDFE.5.std       \\
\hline
\end{tabular}
\end{table}

\subsubsection{Model Interpretability}
Alioth utilizes SHAP, not only for feature selection but also to provide accurate and consistent explanations. SHAP can quantify the magnitude and direction of each feature's impact on the predicted values, based on which we propose an application on attribution analysis. Namely, we view each type of shared resource VMs may contend on as a \textbf{source of interference} (SoI), and try to quantify its contribution to the final degradation.

Suppose we have identified the top-k interference-intensity-correlated low-level metrics $\mathbi{s}_k=\{l_{x_1},l_{x_2},\cdots,l_{x_k}\}$ for each SoI. We now use linear regression to build a simple model of estimating interference intensity level $IIL^i$ of shared resource $i$ :
$ IIL^i=\sum_{j=1}^k \alpha_j^i l_j^i $
and we use Least-Min-Square to optimize the model, and save value of $\omega^i=(\alpha_j^i)_{1\times k}$.
We then enlarge it to $\tilde{\omega}^i$ to the same dimension of input features, following the original input order:
if $x_j$ equals to $l_{x_j}$ and belongs to $s_k$, $\tilde{\omega}^i_j$ is set as $\alpha_{x_j}^i$, if not, we set $\tilde{\omega}^i_j$ as 0.

Suppose prediction for an input $\mathbi{x}$ of $n$-dimension and the corresponding Shapeley value vector $\Phi=\{\phi_1,\phi_2,\cdots,\phi_n\}$ are obtained. For each shared resource $i$, we compute a interfering score $c_i$ based on $\Phi$ and $\omega^i$:
$ c_i=\sum_{j=1}^n \tilde{\omega}^i_j \phi_j $.
We stack the scores as an $n_s$-dim vector $\mathbi{c}=[c_1,c_2,\cdots,c_{n_s}]^\top$, where $n_s$ is the number of SoIs. Finally we normalize $\mathbi{c}$ as
$\tilde{\mathbi{c}}=\frac{\mathbi{c}}{\sum c_i}=[\tilde{c}_1,\tilde{c}_2,\cdots,\tilde{c}_{n_s}]^\top,$
where $\tilde{c}_i$ is the contribution of SoI $i$ made on $\mathbi{x}$'s performance degradation. We present a case study in Sec V-D to demonstrate Alioth's interpretability.

 \begin{figure*}[t]
\setlength{\abovecaptionskip}{0.1cm}
\setlength{\belowcaptionskip}{-0.1cm}
    \centering
    \includegraphics[width=0.9\linewidth]{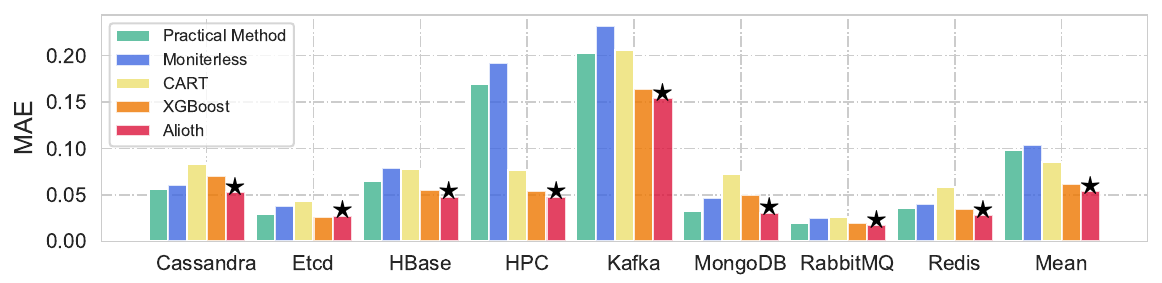}
    \caption{Mean Absolute Error (MAE) of Practical Method,  Moniterless,  CART, XGBoost, and Alioth on different applications. CART is Classification And Regression Tree. XGBoost (eXtreme Gradient Boosting) is a simplification of Alioth with DAE removed. The test and training sets are from the same application and partitioned according to the 8:2 ratio. The best results are marked with an asterisk.}
    \label{fig:metrics}
\end{figure*}

\begin{table*}[htbp]
\setlength{\abovecaptionskip}{0.05cm}
\setlength{\belowcaptionskip}{-0.05cm}
  \centering
  \caption{Online MAE of Practical Method,  Moniterless,  CART, XGBoost, and Alioth on different applications.}\label{tab:online}
    \begin{tabular}{l|rrrrrrrrr}
    \toprule
          & \multicolumn{1}{l}{Cassandra} & \multicolumn{1}{l}{Etcd} & \multicolumn{1}{l}{Hbase} & \multicolumn{1}{l}{HPC} & \multicolumn{1}{l}{Kafka} & \multicolumn{1}{l}{MongoDB} & \multicolumn{1}{l}{RabbitMQ} & \multicolumn{1}{l}{Redis} & \multicolumn{1}{l}{Mean} \\
    \midrule
    Homogeneous Train\&Test  & 0.0508 & 0.0277 & 0.0470 & 0.0542 & 0.1481 & 0.0372 & 0.0207 & 0.0375 & 0.0529 \\
    Oracle DAE+XGBoost & 0.0820 & 0.0654 & 0.0550 & 0.1264 & 0.1720 & 0.0426 & 0.0707 & 0.0567 & 0.0838 \vspace{1pt}\\ 
    \hline 
    Practical Method & 0.1323 & 0.7844 & 0.0677 & 0.1723 & \textbf{0.2124} & 0.2877 & 0.1963 & 0.1587 & 0.2514 \\
    Monitorless & 0.1287 & 0.5673 & 0.0891 & 0.2197 & 0.3144 & 0.2503 & 0.1538 & 0.1479 & 0.2339 \\
    CART  & 0.1264 & 0.9602 & 0.0743 & 0.5741 & 0.4442 & 0.0997 & 0.1776 & 0.1373 & 0.3242 \\
    XGBoost   & 0.1102 & 0.2658 & \textbf{0.0545} & 0.2453 & 0.3037 & 0.0784 & 0.1432 & 0.1440 & 0.1681 \\
    Offline Alioth & 0.1376 & 1.5446 & 0.0883 & 0.7635 & 0.295 & 0.3713 & 0.226 & 0.2512 & 0.4597 \\
    Online Alioth & \textbf{0.1047} & \textbf{0.0839} & 0.0650 & \textbf{0.1628} & 0.2373 & \textbf{0.0487} & \textbf{0.0782} & \textbf{0.0847} & \textbf{0.1082} \\
    \bottomrule
    \end{tabular}%
  \label{tab:addlabel}%
\end{table*}%
\vspace{-0.2em}

\section{Evaluation of Alioth}\label{evaluation}

In this section, we evaluate the performance of Alioth in the following aspects. Firstly, we show the results for classifying applications and identifying unknown applications. Then we evaluate the estimation accuracy by comparing against two representative baseline methods \cite{sc12,monitorless} and two basic ML methods. Thirdly, we show the robustness of Alioth on 0/1 QoS violation of applications under dynamicity. Finally, we discuss the interpretability of Alioth, using a case study to illustrate that Alioth can provide more information to improve the management of cloud systems.

\subsection{Offline Performance of Performance Estimator}\label{sec:7-2}

We first compare our model with two representative baselines on the offline performance. 

 \noindent\textbf{Practical Method}  \cite{sc12} is the first to introduce ML method to this domain. It uses \textit{CfsSubset} in \textit{Weka} to select features 
    and bagged \textit{REPTree} to predict application performance. 
    We re-implement it using \textit{SelectKBest} based on F-value in \textit{scikit-learn} to select features and \emph{bagged decision tree} to predict since they have similar performance. 
    
\noindent\textbf{Monitorless} \cite{monitorless} originally predicts the resource saturation, but 
    it can also be used to predict the QoS degradation with modification. 
    Monitorless adds binary and time-dependent features as feature engineering. 
    It transforms the bytes-values metrics to a logarithmic scale 
    and combines features using multiplication. 
    Random forest is used to select the most relevant features and 
    do the prediction. We follow its original methodology except for data labeling.

Figure~\ref{fig:metrics} compares Alioth with these two representative baseline and two basic ML methods (CART \cite{cart} and XGBoost \cite{xgb}) on all target applications. Alioth outperforms the others, providing the most accurate predictions. Moniterless requires operations to filter and integrate features, while Alioth not only requires no additional operations but also has higher accuracy. The comparison between Alioth and XGBoost shows DAE’s effectiveness in reducing noise and extracting features. It is worth mentioning that Alioth has low makespan and high efficiency. Each training epoch takes an average of 0.45 seconds, and Alioth can converge with less than 200 epochs. The average time of giving inference for a test sample is 1.0048ms on GPU.

\subsection{Online Performance of Performance Estimator}\label{sec:1vr}

What is more important is how these methods perform on unseen and differently distributed data w.r.t. the training data. A key problem is that the real online data is unlabelled, which prevents evaluation. Therefore, we use the experimental scheme described in Sec IV-B 2) to evaluate the online performance. We exclude the application $X$ in the training set, and make the testing set only includes data of $X$. For online Alioth, the testing set acts as the unlabelled target domain data to perform adversarial training.

Table~\ref{tab:online} shows the result of the methods mentioned above. There are two performance baselines, the first is the first row in the table which is performance of Alioth when the training and testing sets are homogeneous (i.e. offline stage performance shown in Fig.~\ref{fig:metrics}). This baseline is used to see how bad the methods deteriorate compared with the offline stage. The second baseline is ``Oracle DAE'', which means to directly use the ground truth feature vectors under no interference rather than predictions of DAE as the input of XGBoost. This baseline can reflect how much effective the designed DADAE is on improving the generalization ability. As the best results are emphasized with bold texts, we can see that Online Alioth achieves most of the best performance and ranks top in terms of average MAE. Online Alioth is also close to ``Oracle DAE'', which demonstrate the effectiveness of DADAE. 

\subsection{Robustness Analysis}\label{sec:7-3}
In real-world scenarios, the criteria for QoS violation are dynamic. Methods like \cite{monitorless} preset a degradation threshold and predict whether an application degrades no longer work when the criteria change dynamically. By taking QoS deterioration ratio as a prediction target, Alioth can dynamically determine application states based on varying thresholds.

\begin{figure}[h]
    \centering
    \includegraphics[width=1\linewidth]{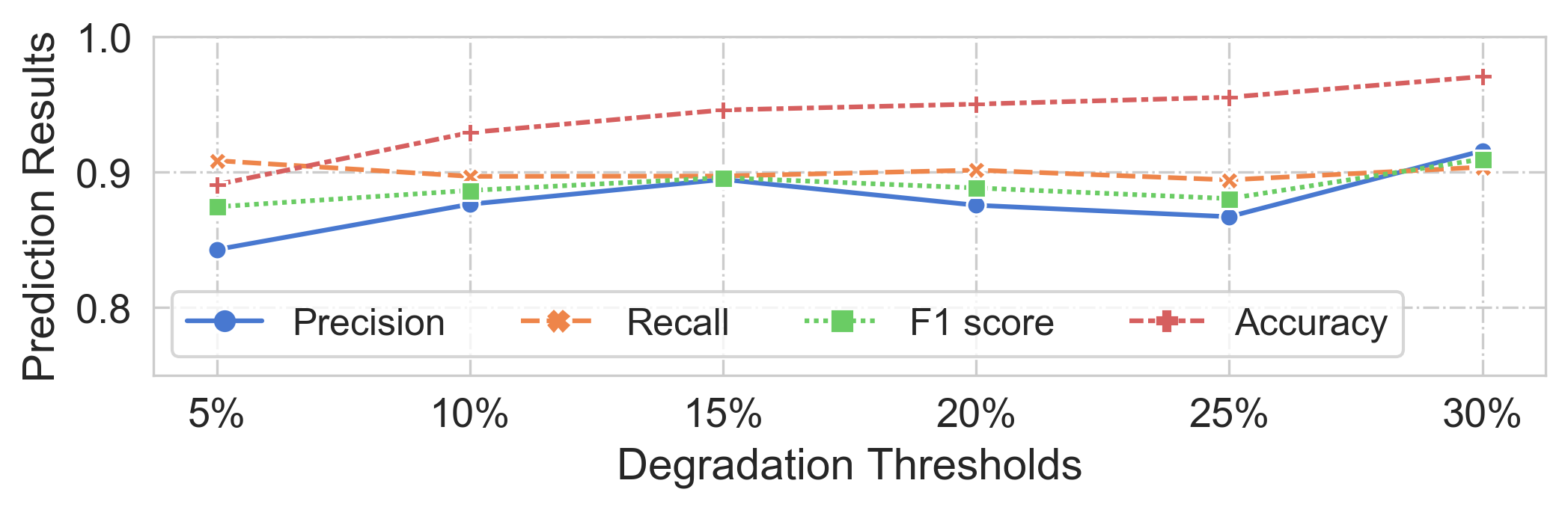}
    \caption{Results of sensitivity analysis. Four typical evaluation metrics of binary classification are considered, and all of their values have very small volatility. The maximum volatility is 0.07, and the minimum is 0.01.}
    \label{fig:sensitivity analysis}
\end{figure}

We set 6 degradation thresholds, from 5$\%$ to 20$\%$ with intervals of 5$\%$, and examine Alioth's average performance in identifying 0/1 QoS violation for all applications. Figure~\ref{fig:sensitivity analysis} shows that Alioth exhibits considerable and robust predictive performance for any degradation threshold. This indicates that Alioth's ability to signal 0/1 QoS violation is not affected by the violation criteria. 

\subsection{Interpretability Analysis}\label{sec:7-4}
 In this part, we use a case study to show how Alioth provides result interpretability with the SHAP explainer.


\begin{figure}[ht]
    \centering
    \includegraphics[width=1\linewidth]{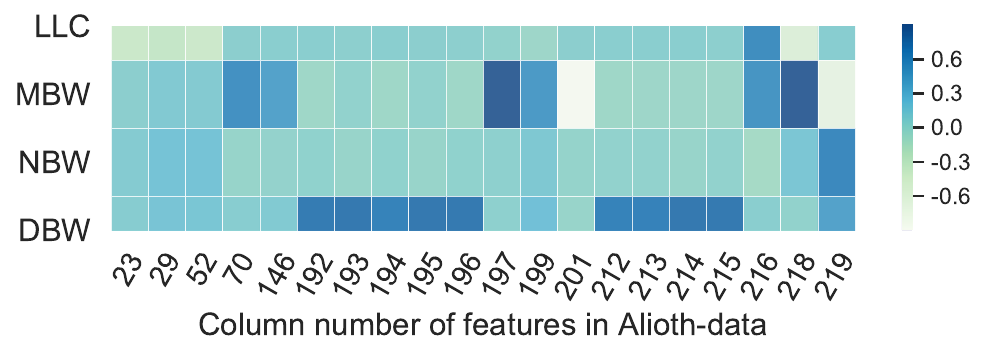}
    \caption{Correlations between 20 top-correlated features and SoI. The feature names are replaced with column numbers in \emph{Alioth-dataset} due to space limit.}
    \label{fig:corr analysis}
\end{figure}

As mentioned in Sec. IV-C 2), we first identify the most relevant low-level metrics with respect to interference intensity levels. Figure~\ref{fig:corr analysis} shows the heat map of the correlation coefficient matrix between the four resources of LLC, MBW, NBW, DBW, and the 20 metrics most relevant to them. Due to space limits, we have to replace these long feature names with their corresponding column numbers in \emph{Alioth-dataset}. 
For example, the feature most relevant to DBW is No.194 \emph{bread\_S}, which describes the total block reads on the host and complies with intuition. 


 \begin{figure}[ht]
    \centering
    \subcaptionbox{\label{fig:find_SoI}}[0.43\linewidth]{
    \includegraphics[width=\linewidth]{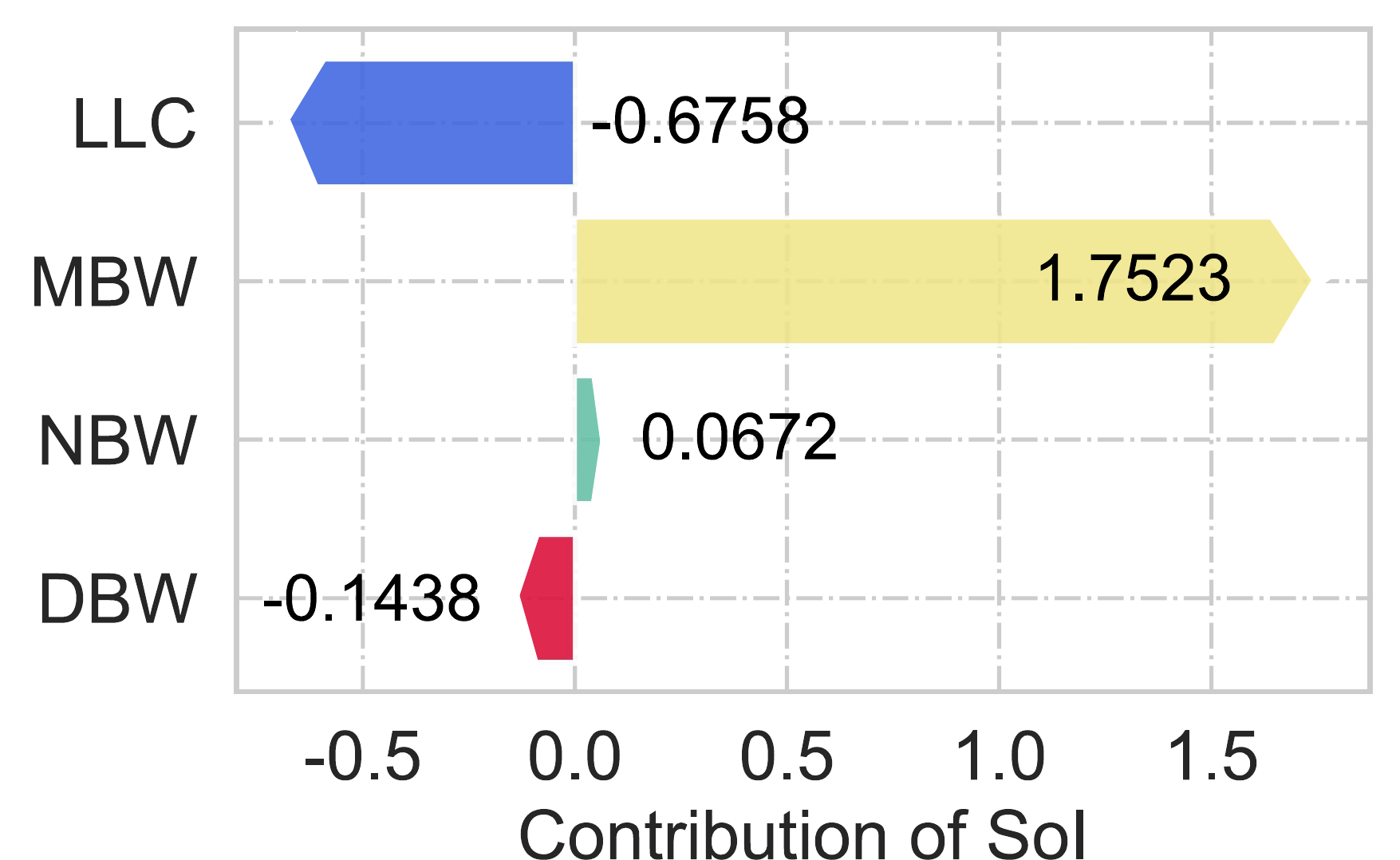}
    }
    \subcaptionbox{\label{fig:Attribution analysis}}[0.53\linewidth]{
    \includegraphics[width=\linewidth]{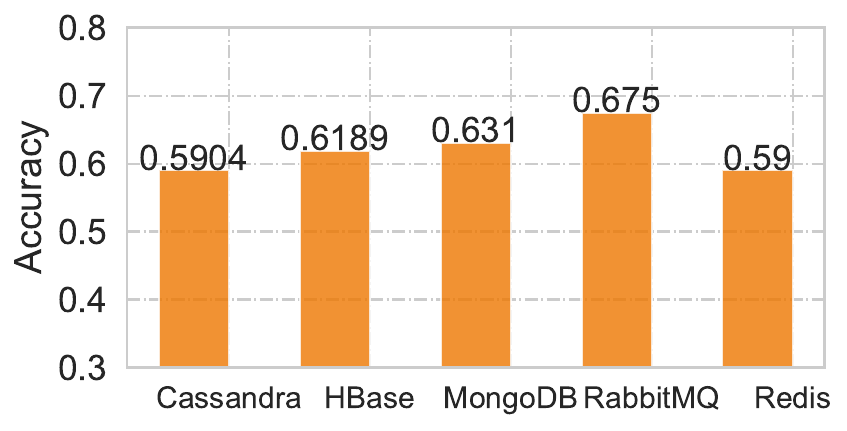}
    }
    \setlength{\abovecaptionskip}{0.2cm}
    \caption{Results of attribution analysis. (a) shows the contribution of SoI for one test sample. (b) presents the accuracy for 5 applications.}
\end{figure}

We then take RabbitMQ as an example to demonstrate how to use Alioth to perform attribution analysis. 
For a testing sample, if Alioth estimates that there is a QoS violation (violation threshold set at \textbf{5$\%$}), we calculate a 1$\times$4 vector $\mathbi{c}$. Each value of the vector corresponds to the contribution of  LLC, MBW, NBW, and DBW to the estimated result. The resource with the highest value $\tilde{c}_i$ can be seen as the major reason for QoS violation. For example, if Alioth estimates $\hat{D}_{v,t}=$\textbf{0.1008}, with the contribution of each SoI shown on the left of Figure~\ref{fig:find_SoI},  we can infer that \textbf{MBW} is the main SoI.

We further use the interference type information in \emph{Alioth-dataset} to examine the accuracy of this attribution analysis procedure. The top-1 accuracy results of the attribution analysis for several applications are shown in Figure~\ref{fig:Attribution analysis}. The mean accuracy is 0.62, which indicates that Alioth can help cloud platform operators identify the main sources of interference for application performance degradation and provide evidence to support subsequent scheduling operations.

\section{Conclusions}\label{sec:conclusion}

To conclude, in this paper, we propose Alioth, a novel ML based framework as an interference-aware performance monitor of multi-tenant applications in public clouds. 
Alioth relies solely on low-level metrics that can be easily acquired, and well suits various workload intensity and interference types.
To satisfy the data needs of ML methods, we first elaborate interference generators and conduct a set of comprehensive co-location experiments to make \emph{Alioth-dataset} that reflects the real-world public clouds. 
Then we design the  framework of Alioth with techniques like DAE and DANN to balance model accuracy and generalization.  
We further enhance the model interpretability and automate feature selection via devising a SHAP based explainer.
Our comparative experimental results demonstrate the effectiveness of Alioth, outperforming baseline methods on estimating precision. 
Finally, we present how to perform attribution analysis with Alioth's explainable results to provide more insights for cloud system operators.

There are several issues remaining for future work, including but not limited to: the influence of CPU overcommitting, other categories and QoS metrics of cloud services, scheduling algorithms based on Alioth's results, identifying the interference intensity generated by the applications, etc.. We are also about to deploy our framework in production clusters of Alibaba Cloud to examine Alioth with real-world data.

 \section*{Acknowledgment}
  This work was supported by the National Key R\&D Program of China [2020YFB1707903]; the National Natural Science Foundation of China [62272302, 62172276], Shanghai Municipal Science and Technology Major Project [2021SHZDZX0102], and the Alibaba Cloud [TC20201127009,TC20220718012]. We thank Shuodian Yu, Yiming Liu, Zifeng Liu from Shanghai Jiao Tong University, Xuqi Zhu, Shuming Jing, Zhiyuan Cai from Alibaba Cloud for help building \emph{Alioth-dataset} and enhancing this work. We also thank the anonymous reviewers for their feedback.

\bibliographystyle{IEEEtran}
\bibliography{conference_101719}

\end{document}